\documentstyle[aas2pp4]{article}

\def\lsim{\lower.5ex\hbox{$\; \buildrel < \over \sim \;$}}
\def\gsim{\lower.5ex\hbox{$\; \buildrel > \over \sim \;$}} 
\def\lax    {\ifmmode{_<\atop^{\sim}}\else{${_<\atop^{\sim}}$}\fi}
\def\gax    {\ifmmode{_>\atop^{\sim}}\else{${_>\atop^{\sim}}$}\fi}

\def\gtorder{\mathrel{\raise.3ex\hbox{$>$}\mkern-14mu
	     \lower0.6ex\hbox{$\sim$}}}
\def\ltorder{\mathrel{\raise.3ex\hbox{$<$}\mkern-14mu
	     \lower0.6ex\hbox{$\sim$}}}

\begin{document}

\title{Modeling the Hard X-Ray Lag and \\
       Energy Spectrum of Cyg X-1}

\author{ Xin-Min Hua\altaffilmark{1,2}, Demosthenes Kazanas\altaffilmark{1}, 
and Wei Cui\altaffilmark{3}} 

\altaffiltext{1}{LHEA, NASA/GSFC Code 661, Greenbelt, MD 20771}
\altaffiltext{2}{Universities Space Research Association}
\altaffiltext{3}{Center for Space Research, MIT, Cambridge, MA 02139}

\font\rom=cmr10
\centerline{\rom submitted to Astrophys. J., June 24, 1997}

\begin{abstract}
In an effort to model the observed energy spectrum of Cygnus X-1 as 
well as its hard X-ray lag by Comptonization in inhomogeneous  clouds 
of hot electrons with spherical geometry and various radial 
density profiles we discovered that: 1) Plasma clouds with different 
density profiles will lead to different Comptonization energy spectra 
even though they have the same optical depth and temperature. On the other 
hand, clouds with different optical depths can produce the same energy 
spectra as long as their radial density distributions are properly chosen. 
Thus by fitting the energy spectrum alone, it is not possible 
to uniquely determine 
the optical depth of the Comptonization cloud, let alone its density structure.
2) The phase or time difference as a function of Fourier frequency or
period for the X-rays in two energy bands is sensitive to the radial density 
distribution of the scattering cloud. Comptonization in plasma clouds 
with non-uniform density profiles can account for the long standing 
puzzle of the frequency-dependent hard X-ray lags of Cygnus X-1 and 
other sources. Thus simultaneously fitting the observed spectral and 
temporal X-ray properties will allow us to probe the density structure of 
the Comptonizing atmosphere and thereby the dynamics of mass accretion
onto the compact object.
\end{abstract}


\section{Introduction} 
Much of our understanding of the physics and dynamics of accretion
in X-ray binaries has been achieved mainly through the modeling and
interpretation of their high energy photon spectra. The high
energy emissions are believed to originate in the vicinity of the 
compact object and is considered to be due to the Comptonization of 
soft photons by the high temperature ($T_e \sim 10^9$ K)
electrons heated by the dissipation of the accretion kinetic energy.
This process has been studied extensively (e.g. Sunyaev \& Titarchuk
1980; Hua \& Titarchuk 1995) and model photon spectra in analytic forms
have been used, quite successfully, to fit the observed spectra of 
accreting sources, in the energy range of a few to a few hundred keV. 
These fits can then provide the values of the Thomson optical depth 
$\tau_0$ and the temperature $T_e$ of the hot electrons. While the 
parameters so obtained are widely accepted as the true indication of 
the physical conditions prevailing in the region under consideration, 
one should keep in mind that the analytical formulas  derived are based 
on several approximations. Hua \& Titarchuk (1995) showed that in order 
for the analytical formulas to be valid, the optical depth $\tau_0$ and 
the electron temperature $T_e$ must be within certain limits and the 
initial distribution of source photons must satisfy certain requirements. 
Furthermore, the analytical treatment can only by applied to the 
simplest geometries, typically spheres or disks with uniform temperature 
and density. Obviously, these requirements are not realistic in most 
situations. 

For example, Skibo \& Dermer (1996) and Ling et al. (1997) have shown 
that for some of the observed energy spectra, two zone models of 
different temperatures can significantly improve the energy spectral 
fits. Along the same lines, Kazanas et al. (1997, hereafter KHT), 
motivated by considerations of the accretion dynamics and the variability 
of the observed light curves as manifest in the form of power spectra 
(Miyamoto et al. 1992, Cui et al. 1997b), introduced a model consisting 
of an extended Comptonizing atmosphere in spherical geometry, with 
density profiles $r^{-1}$ or $r^{-3/2}$, where $r$ is the distance from 
the center of the spherical clouds. Hua et al. (1997, hereafter HKT)
further showed that the hard X-ray time lags and the sources' coherence 
functions, derived from  observations of X-ray binaries by Ginga 
(Miyamoto et al. 1991, Vaughan \& Nowak, 1997), and also by more recent 
observations by RXTE  (Cui et al. 1997b) are consistent with the 
inhomogeneous models introduced by KHT and inconsistent with the uniform 
ones. 

These more realistic models have more adjustable parameters and  
before they can be used to interpret the observational data, the  
question arises as to how the energy spectra resulting from 
such models depend on these extra parameters. An additional question is whether
there exist any observational tests which can provide  information 
concerning the inhomogeneity present in these more general models. The 
purpose of the present paper is to provide answers to these questions.

In \S 2, we show how the radial density profiles of  scattering clouds
with a given optical depth and temperature can affect the emergent 
energy spectra. In \S 3, we  provide an example of spectral fitting of the 
Cygnus X-1 data obtained in combination by CGRO/BATSE and RXTE/PCA 
using model spectra corresponding to configurations of different density 
profiles. In \S 4, we show that X-ray time lags corresponding to these 
different models are sufficiently different to serve as a discriminator of 
these models, thus giving preference to a particular class of inhomogeneous 
model. In \S 5, we present a simple analytic model which provides a 
transparent account of the numerical results. Finally, in \S 6 we provide 
a summary and conclusions of our results and we discuss the implication 
of these models for the interpretation of observational data.

\section{The Effects of The Density Profile on The Energy Spectrum}

In the present study we concentrate on the Comptonization process in 
hot electron clouds with  density profiles of the form 
$$n(r) = \cases  {n_i &for $r \le r_1$ \cr n_1 (r_1/r)^{p} &
for $r_2 > r > r_1$ \cr} \eqno(1)$$
\noindent 
where $p$ is a free parameter; $r$ is the radial distance from
the center of a spherical cloud; $r_1$ and $r_2$ are radii of the
inner and outer edges of the extended non-uniform ``atmosphere''
respectively. The central core is assumed to be uniform and its 
density is $n_i$, which may not necessarily equal to $n_1$.

For illustration purposes, we have calculated the energy spectra emerging 
from clouds with the same total optical depth $\tau_0 = 1$ and 
electron temperature $kT_e = 100$ keV but different density profiles, 
namely $p=3/2$, 1 and 0 in Eq. (1). The first two profiles represent 
the dynamically more plausible free-fall ($p=3/2$) density distribution 
such as implied by the advection-dominated accretion model suggested 
by Narayan \& Yi (1994) and the accretion induced by the sort of radiative 
viscosity ($p=1$) suggested by KHT. For the uniform profile ($p=0$), 
traditionally used in spectral fitting because of its mathematical 
simplicity, we further assume that it has an electron density of the
order $10^{16-17}$ cm$^{-3}$ and a size of the order $\sim 
10^{-3}$ light second, typical of the values thought appropriate for 
the conditions in the vicinity of the compact object in galactic 
accreting sources. For the other 
two profiles, both central cores have radius $r_1 = 1.0\times 10^{-3}$ 
light seconds. In the absence of detailed knowledge of this innermost 
part of accreting cloud, we simply assume it has uniform density 
$n_i$ such that its Thomson optical depth $\approx \tau_0/3$. 

\begin{figure}[ht]
\plotone{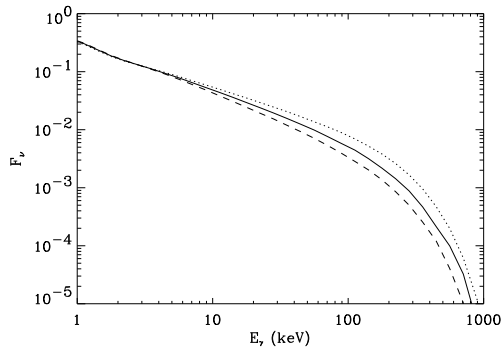}
\caption{\baselineskip=10pt
Energy spectra resulting from Comptonization in clouds with different
density profiles. The clouds have the same optical depth $\tau_0 =1$ and
plasma temperature $kT_e = 100$ keV. The spectra are from clouds with
uniform (dotted), $p=1$ (solid) and $p=3/2$ (dashed curve) density
profiles respectively.  }
\end{figure}

Model energy spectra obtained using the Monte Carlo code of Hua (1997),
developed to deal with photon propagation and Compton scatterings in 
inhomogeneous media, for conditions appropriate for these 
density profiles are shown in Figure 1. It is seen that although the 
emissions are from clouds with the same optical depth and plasma 
temperature, their spectra are significantly different from each other. 
In general, for the same $\tau_0$, $T_e$, $r_1$, $n_1$ and $n_i$ in
Eq. (1), the greater the density gradient represented by $p$, the 
softer the emergent photon energy spectrum for a given total Thomson 
depth $\tau_0$. This is because photons propagating in media with a density 
gradient are in an anisotropic environment -- the outward scatterings have 
longer mean free path than the inward ones. Consequently, clouds
with greater density gradient are more difficult to trap the photons,
leading to softer energy spectra.

\section {Modeling the Energy Spectrum \\ of Cyg X-1}

The above discussion indicates that the optical depth and plasma 
temperature alone are not sufficient to determine the spectrum due to the 
Comptonization process. Clouds with the same
optical depth and temperature can lead to different Comptonization
spectra if they have different density distributions. On the other
hand, clouds with different optical depths or temperature can lead to
the same photon spectrum as long as their density profiles are properly 
chosen. This can be clearly seen in the following example of spectral 
fitting in which three photon spectra resulting from clouds with 
the same temperature but different optical depths and density profiles 
fit equally well a given  set of observational data. The data are taken 
from Table 2 of Ling et al. (1997), which tabulates the spectrum of 
the blackhole candidate Cyg X-1 at its $\gamma_0$ state observed in late 1993. 
Ling et al. (1997), using a Monte Carlo simulation of Comptonization 
in uniform density clouds, obtained a best-fit model corresponding to 
a spherical cloud with optical depth $\tau_0 = 0.435$ and temperature 
$kT_e = 107.7$ keV.

The spectrum was obtained by Comptonization of seed photons from a black body
distribution of temperature $kT_e= 0.5$ keV. The source of photons was 
considered to lie outside the hot electron cloud and to be
injected radially inward as suggested by Skibo \& Dermer (1995).
Here we re-fit the same data using 
a slightly different model with $\tau_0 = 0.5$, $kT_e = 100$ keV 
and source photons at $kT_0 = 0.2$ keV injected isotropically at the 
cloud center. The cloud is assumed to be uniform as in the previous 
fitting. The spectrum resulting from such a model is displayed in Figure 2
(dotted curve) together with the data points. The reduced $\chi^2$ value 
is 10.3 with 11 degrees of freedom, not as low as the one 
obtained by Ling et al. (1997) but still perfectly acceptable.

\begin{figure}[ht]
\plotone{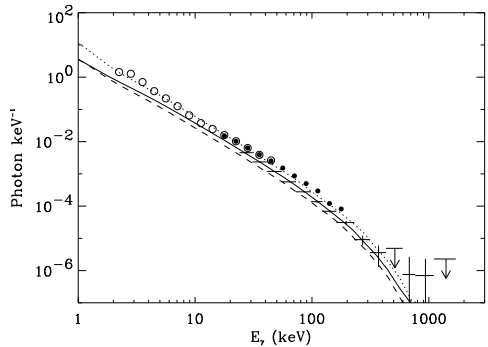}
\caption{\baselineskip=10pt
Three calculated energy spectra which fit equally well the Cyg X-1
data (crosses) in its $\gamma_0$ state observed by CGRO/BATSE in 1994
(Ling et al. 1997). These spectra result from Comptonization in
clouds with the same temperature but different optical depths
and density profiles. The dotted and dashed curves are slightly
displaced to separate the otherwise nearly identical curves.
Also plotted are RXTE/PCA (circles) and HEXTE (dots) data from the
same source observed during its high state in 1996 (Cui et al. 1997a).}
\end{figure}

In the same figure, we also display the Monte Carlo calculated spectra
resulting from two other models which have the same plasma temperature 
$kT_e = 100$ keV but different optical depths $\tau_0$ and non-uniform
density profiles. One of them is a cloud with density profile parameter
$p=1$ and $n_1 = 4.35\times 10^{16}$ cm$^{-3}$. Its inner core has a 
uniform density and a radius  $r_1 = 10^{-4}$ light second with a 
Thomson optical depth $\tau_1 = 0.2$. The total optical depth of the 
cloud is 1 and the energy spectrum resulting from Comptonization 
in such a cloud is displayed in Figure 2 as the solid curve. The other 
model is a cloud with $p=3/2$ and $n_1 = 1.594 \times 10^{17}$ cm$^{-3}$. 
Its uniform inner core has the same radius but  a Thomson optical    
depth $\tau_1 = 0.07$. The total optical depth of the cloud is 0.7 and
the energy spectrum from such a cloud is displayed in Figure 2 as the 
dashed curve. It is seen that over the energy range from 20 to 200 keV, 
these spectra are almost identical to the dotted curve. In fact, 
the $\chi^2$ values are 7.8 and 8.1 respectively (with 8 degrees of freedom). 
In other words, these two models are as good as the uniform one
insofar as the spectral fits are concerned.

Since these spectra were produced by a Monte Carlo calculation, a few words 
about its statistical errors are in order. Each of the spectra was obtained 
by following $8 \times 10^6$ photons. By generating several such spectra 
under the same conditions but with a different initial random number seeds 
and by fitting the spectra to the same data, we were able to estimate the 
statistical uncertainty of the calculation. It was found that the 
standard deviation of the $\chi^2$ value introduced by the Monte Carlo 
statistics is less than 2, which is small
compared to the $\chi^2$ values obtained. This justifies the conclusion 
about the goodness of the fits. 

In Fig. 2 we also plotted the PCA (circles) and HEXTE (dots)
data covering an energy range $2-200$ keV from RXTE observations of 
Cyg X-1 in its 1996 high state (Cui et al. 1997a). By definition, the 
soft X-ray ($\lsim 10$ keV) flux is high in the high state, 
but the hard flux is low. Such a distinct anti-correlation 
between the soft and hard bands was confirmed by the simultaneous monitoring 
of Cyg X-1 with the ASM/RXTE and BATSE (Cui et al. 1997a, Zhang et al.
1996). Cyg X-1 in the $\gamma_0$ state, as defined by Ling et al. (1997), 
showed a similar BATSE flux. So it might have been in the high state as 
well in 1994. Unfortunately, no simultaneous soft X-ray coverage 
was available then to confirm it. On the other hand, the remarkable 
similarity in the observed X-ray spectral shape, as is evident from the
figure, between these two observations 2.5 year apart also 
seems to suggest a common origin. If so, the results would imply 
that the high state spectrum of Cyg X-1 (and perhaps the high state 
itself) is quite stable, except for a slight shift in the normalization 
(part of which can be attributed to the uncertainty in the cross 
calibration between the RXTE instruments and BATSE).  

\section {Hard X-Ray Time Lag of Cyg X-1}

Although  clouds with different density profiles can produce the 
same energy spectrum, at least over a given energy range, as long as 
they have properly chosen $\tau_0$ and $T_e$, KHT and HKT have shown that 
the time-variation properties of the  sources are sensitive 
to the density distribution of the Comptonizing atmosphere and therefore 
they can be used to distinguish the models of
different density profiles. In fact, the inhomogeneous density profile 
models were conceived by KHT in order to explain the 
observed temporal properties such as power spectral density (PSD) and 
hard X-ray phase or time lags.  In Figure 3, we show the 
time lags of the X-ray emission as a function of Fourier period (see 
e.g. van der Klis et al. 1987) resulting from the three model clouds 
which were used to  produce the energy spectra shown in Figure 2. 
These were obtained using the Monte Carlo code by collecting the 
escaping photons according to their arrival time to the observer 
as well as their energy. The photons were collected in the two energy 
bands $15.8 - 24.4$ and $1.2 - 5.75$ keV in order to be directly 
compared to the observational data obtained by Ginga (Miyamoto et al. 
1988). In each energy band, the photons were collected into 4096 bins 
over 16 seconds, each $1/256$ seconds in length. The light curves 
so obtained were then used to calculate the time lags of the emission 
in the $15.8 - 24.4$ keV band with respect to that in the $1.2 - 5.75$ keV
band. The time lag resulting from the uniform cloud (dotted curve) is 
obtained by further assuming the cloud has a density 
$n = 10^{16}$ cm$^{-3}$. It is seen that the 
lag decreases below the period $\sim 0.005$ second but becomes constant 
$\approx 2$ ms above it. This is because for emission from such a cloud, 
the hard X-ray lags are due to scatterings in a region with a mean free 
time of the order $0.3/n\sigma_T c \simeq 1.5$ ms (Hua \& Titarchuk, 1996). 
The magnitude of the time lag reflects this characteristic time. 

\begin{figure}[ht]
\plotone{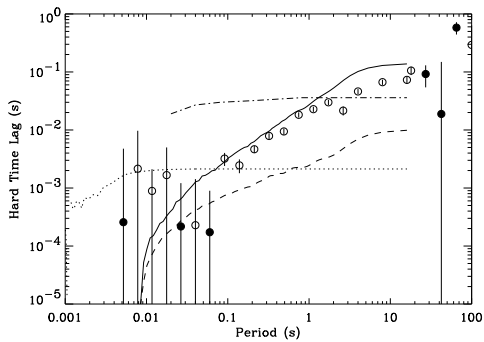}
\caption{\baselineskip=10pt
The time lags of hard X-rays ($15.8 - 24.4$ keV) with respect to soft ones
($1.2 -5.7$ keV) resulting from Comptonization in the same clouds that
produce the energy spectra shown in Figure 2. The dotted, solid and
dashed curves, as in Figure 2, represent the density profiles $p=0, 1$ 
and 3/2 respectively. The time lag between the same energy bands 
based on  Ginga data from Cyg X-1 (Miyamoto et al. 1988) are also plotted
in the figure.}
\end{figure}

On the other hand, the time lag resulting from the cloud with $p=1$ 
density profile (solid curve) has a linear dependence on the Fourier 
period ranging from 0.03 second up to $\sim 3$ second. There is a 
cutoff at the periods below the range due to the finite time resolution 
of our calculation. At periods $P \gsim 4$ seconds the curve levels
off, indicating that the time lag of hard X-ray reaches its maximum of 
$\sim 0.1$ second. This is because the lags in this case are due
to scatterings in a region with densities ranging from $\sim 10^{16}$ to
$10^{12}$ cm$^{-3}$. In addition, the probabilities for photons to 
scatter in each decade of density are equal (KHT). Correspondingly, the 
scattering mean free time spans a range of four orders of magnitude.
As a result, the time lags also span a range of four orders of magnitude, 
and the level-off period roughly indicates the size of the cloud, which
is 1 light second in this case. The time lag resulting from the cloud 
with $p=3/2$ density profile is also plotted in Figure 3 (dashed curve). 
It is seen that dependence of lag on period is weaker and the maximum 
time lag is only $\sim 10^{-2}$ second, indicating that although the 
cloud extends to a radius of $\sim 1$ light second, because of the steep density
gradient, there is virtually no photon scattering beyond a  radius 
$\sim 10^{-2}$ light second. A calculation based on the density profile 
of Eq. (1) indicates that the optical depth of the cloud for 
$r > 10^{-2}$ light second is $\sim 7.5\%$ of the total depth for $p=3/2$, 
while in the case of $p=1$, the corresponding percentage is $40\%$.

Also plotted in Figure 3 is the time lag observed
from Cyg X-1 between the same energy bands by Ginga (Miyamoto et al.
1988). It is seen that, qualitatively, the nearly linear dependence of
time lag on Fourier period is best fitted by the curve
representing the cloud with $p=1$ density profile. The curve
for the cloud with $p=3/2$ density profile, because of its weaker
dependence on the Fourier period, obviously does not provide as good a fit 
to the data and could be considered excluded for this specific
data set. It is also obvious that a
uniform cloud model can be excluded, a conclusion reached also by
Miyamoto et al (1988) by comparing the time lag obtained based on the 
analytic formula of Payne (1980) for a uniform cloud. The latter time lag
corresponding to a density $n = 10^{16}$ cm$^{-3}$ is plotted in the 
figure as a dash-dotted curve. The difference between the analytic curve 
and Monte Carlo result (over one order of magnitude) is mainly 
due to the relativistic electron temperature and small optical depth 
as explained in Hua \& Titarchuk (1996). 

However, the  Ginga data presented in fig. 3 were obtained in August 1987, 
when Cyg X-1 was in its low state. Therefore, it may be considered 
inappropriate to compare them with the calculated time lags based on 
the cloud configuration used to fit the energy spectrum of the 1993 
observation, when the source was in its high (soft) state. On the other 
hand, as pointed out in \S 3, the RXTE observations of Cyg X-1 in 1996
(Cui et al. 1997a) not only provide data in its high (soft) 
state but also data whose spectrum is similar to that of the 1993 
observation. The time lags of the RXTE observation were reported in
Cui et al. (1997b). As an example, here we use their results from
observation \#6 for comparison, when the source was in its high (soft)
state while the other two reported (\# 3 and \# 15) were during the 
transitions. The Fourier period dependence of the time lag of the energy band 
$13 - 60$ keV with respect to $2 - 6.5$ keV is plotted in Figure 4
with the data points re-grouped into logarithmically uniform bins in
Fourier period. Again, the linear dependence of the lag on 
the Fourier period in the range $\sim 0.03 - 3$ second is evident. 
The lags between the same energy bands calculated for the three 
model clouds described above are also plotted in the figure. The shapes
of the curves are similar to those in figure 3, but the magnitudes of
the lag are slightly smaller, clearly because of the smaller gap between 
the reference energy bands in this case. Obviously, the cloud with $p=1$ 
is again favored by the observation.

\begin{figure}[ht]
\plotone{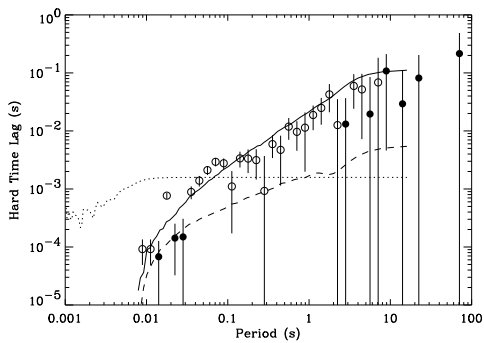}
\caption{\baselineskip=10pt
The time lags of hard X-rays ($13 - 60$ keV) with respect to soft ones 
($2 -6.5$ keV) resulting from Comptonization in the same clouds that
produce the energy spectra shown in Figure 2. The dotted, solid and
dashed curves, as in Figure 2, represent the density profiles $p=0, 1$ 
and 3/2 respectively. The time lag between the same energy bands 
based on RXTE data from Cyg X-1 (Cui et al. 1997b) are also plotted in 
the figure.}
\end{figure}

In addition to these Cyg X-1 data, Ginga data from GX399-4 during its
very high state (Miyamoto et al. 1991) also show a nearly linear 
dependence of the hard X-ray time lags on the Fourier period over the 
range from $\sim 0.1$ to $\gsim 10$ seconds. The lags at low frequencies could 
be as large as $\sim 1$ second. More recently, CGRO/OSSE data from the source
GRO J0422+32 (Grove et al 1997) show a similar dependence for photons in the
energy band  75 $-$ 175 keV with respect to those in the band  
$35 - 60$ keV over the Fourier frequency range $0.03 - \sim 10$ Hz. The time 
lag reaches $\sim 0.3$ second and levels off at $\sim 0.03$ Hz. Therefore
we have reason to believe that the nearly linear time lag dependence on
Fourier period is a common phenomenon shared by many accreting compact 
objects.

The above comparisons for the Cyg X-1 data, qualitative as they are, show 
that by detailed fitting to the observed time lag as a function of frequency 
or period, one can determine the density structure of the Comptonizing 
cloud. More importantly, they show that, if in addition simultaneous 
spectral data are available, the combined analysis of spectral and 
temporal data from the same source can uniquely determine the physical 
size, density profile as well as optical depth, temperature of the 
plasma cloud responsible for the X-ray emissions. We are 
planning to do more detailed data fitting to the energy spectra, time 
lag and other properties observed simultaneously from Cyg X-1 and other 
sources. 

\section {Discussion}

The difference of the time lag curves resulting from Comptonization in
clouds with uniform and non-uniform density profiles, in particular the
one with $p=1$, can be better 
understood if we examine the detailed form of their corresponding high 
energy light curves in response to an instantaneous input of soft photons 
at their center. In KHT, it was shown that the light curve resulting 
from Comptonization in a uniform cloud is exponential in time while 
that from a cloud with $p=1$ density profile has a power law shape, with 
power index close to one which decreases with increasing total Thomson 
depth and photon energy. In both cases there is a cutoff corresponding 
to the time scale characteristic of the photon escape time from the system 
$\beta$. These light curves can be approximated very well analytically
by the Gamma distribution function   
$$g(t) = \cases{t^{\alpha -1}e^{-t/\beta}, &if $t\ge 0$; \cr
                0, &otherwise, \cr } \eqno(2)$$

\noindent
where $t$ is time; $\alpha > 0$ and $\beta > 0$ are parameters determining 
the shape of the light curves. 

\begin{figure}[ht]
\vbox to2.0in{\rule{0pt}{2.0in}}
\includegraphics{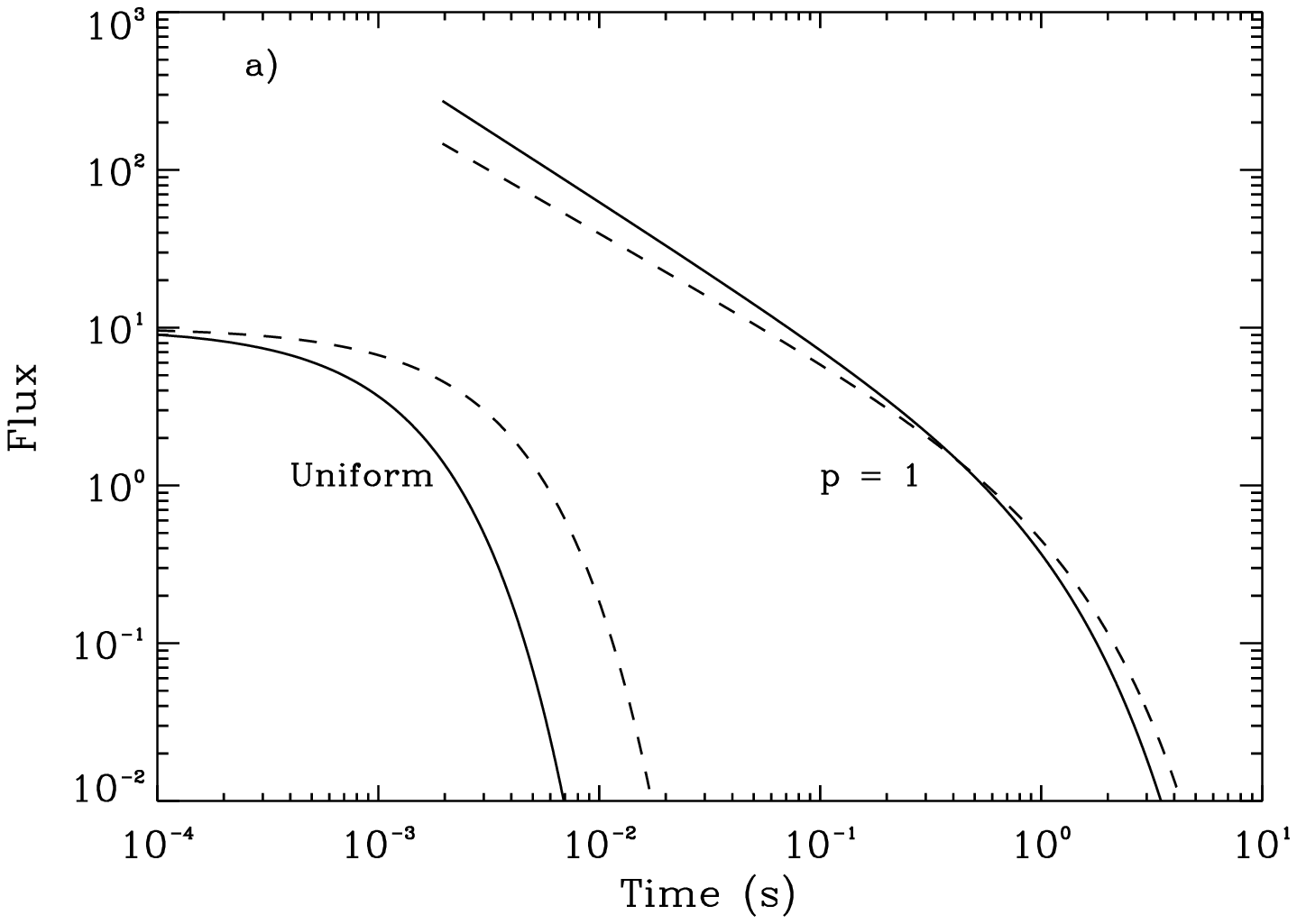}
\vskip -0.2 truein
\vbox to2.0in{\rule{0pt}{2.0in}}
\includegraphics{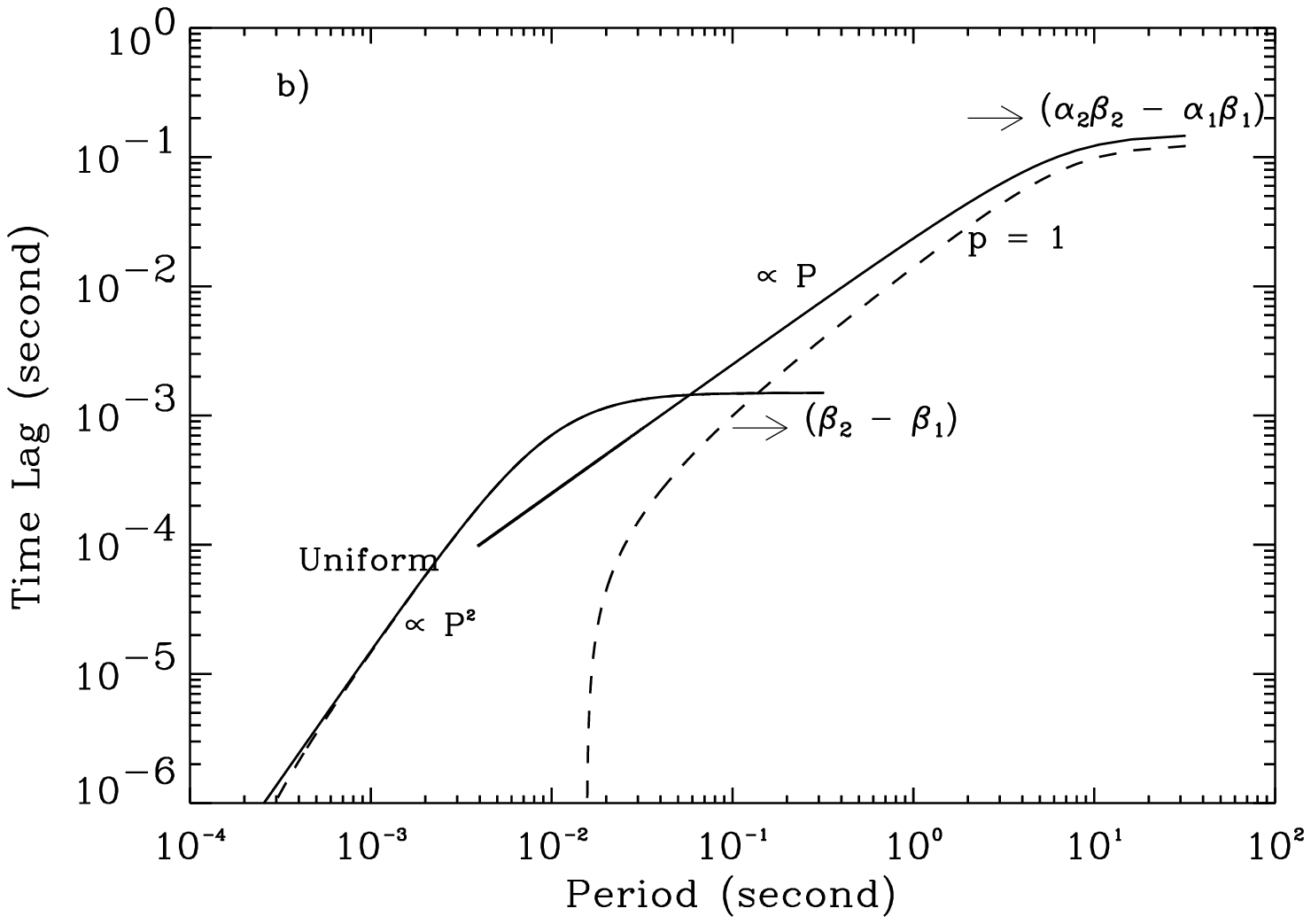}
\caption{\baselineskip=10pt
(a) Two pairs of light curves corresponding to two sets of values of
$\alpha$ and $\beta$ in Eq. (2). The pair with $p= 1$ 
represents the light curves from a cloud with $1/r$ density profile,
the other pair from a uniform one. The solid curves correspond to 
the light curves in the lower energy bands. (b) The time lag based on 
Eq. (5) between the two curves in each pair in (a). The dashed curves
are times lags obtained by numerical Fourier transformation with finite
time resolution.}
\end{figure}

The light curve  from a uniform electron cloud corresponds to $\alpha = 1$ 
so that the Gamma distribution function reduces to an exponential. 
Because the cloud in this case is assumed to be confined to the 
vicinity of the compact object, the corresponding value for $\beta$ is 
of order of $10^{-3}$ sec. The cloud with the $p=1$ density profile under
consideration is much more extended spatially, with its outer radius at 
$\sim 1$ light second, corresponding to $\beta \sim 1$ second, and with a 
value for $\alpha$ small compared to 1. The Fourier transformation 
of $g(t)$ is
$$G(\omega) = \displaystyle{{\Gamma(\alpha) \beta^{\alpha}}\over
              {\sqrt{2\pi}}}(1+\beta^2\omega^2)^{-\alpha/2}
              e^{i\alpha\theta}. \eqno(3)$$

\noindent 
where $\Gamma(x)$ is the Gamma function; $\omega = 2\pi/P$ is the Fourier
frequency and $\theta$ is the phase angle that we are interested in and 

$$\tan{\theta} = \beta\omega. \eqno(4)$$

If we consider two light curves in two energy bands distinguished by
the different values of $\alpha$ and $\beta$, say $\alpha_1$, $\beta_1$ and
$\alpha_2$, $\beta_2$, the time lag between them will be given by their 
phase lag $\theta$ divided by the corresponding Fourier frequency $\omega$, 
i.e.

$$\delta t = \displaystyle{1\over \omega}~[\alpha_2\arctan(\beta_2\omega) - 
\alpha_1\arctan(\beta_1\omega)] \eqno(5)$$

\noindent
For large periods $P$ or $\beta_1 \omega,~\beta_2 \omega \ll 1$, $\delta t$
approaches the constant $\alpha_2\beta_2 - \alpha_1\beta_1$. On the other
hand, for small $P$, $\delta t \simeq (\alpha_2-\alpha_1)P/4$. The transition
from latter to the former occurs at $\beta \omega \sim 1$ or $P \sim
2 \pi \beta$. In our model for $p=1$ density profile, $\beta_1$ and 
$\beta_2 \sim 1$ second. This explains what we see in Figures 3 and 4, namely 
for $P \lax $ a few seconds, $\delta t \propto P$ and the curve levels off 
for large $P$. On the other hand, for the case of uniform cloud, 
$\alpha_1=\alpha_2=1$; $\beta_1$ and $\beta_2 \sim 1$ ms (\S 4).  
As a result, for large periods $P$, $\delta t$ approaches the 
constant $\beta_2 - \beta_1$, which is of the the order 1 millisecond.
For small $P$, $\delta t \propto (1/\beta_1 -1/\beta_2)P^2$.
The transition occurs at $P \sim 2 \pi \beta \simeq 0.006$ second. 

In Figure 5a, we plot two pairs of light curves, one with $\alpha_1 = 0.1$
$\alpha_2 = 0.2$, $\beta_1 = 1$ and $\beta_2 =1.25$, the other with
$\alpha_1 = \alpha_2 = 1.0$, $\beta_1 = 0.001$ and $\beta_2 =0.0025$. 
The former represents the light curves resulting from the cloud with
$p=1$ density profile while the latter from a uniform one. The time lags
between these two pairs of light curves are presented in Figure 5b.
It is seen that the light curve parameters $\alpha$ and
$\beta$ determine the shape of the time lag curve: For $\alpha = 1$,
or pure exponential light curves, the time lag is proportional to $P^2$ 
for small Fourier period $P$ and turns to constant for large $P$. For
$0 < \alpha < 1$, the time lag is proportional to $P$ for small $P$
and turns to constant for for large $P$. In both cases, the level-off
point is $P \approx 2\pi\beta$. However, the time lag resulting from 
the exponential light curves has no portion linear in $P$. Thus the 
existence of a linear portion in the time lag curves obtained from
observations clearly favors the power law light curve. From KHT, we
know that power law light curve is a signature of the non-uniform 
density distribution of the source and the values of $\alpha$ and
$\beta$ are closely related to the physical size and density
distribution of the source cloud. 

Thus, under the assumptions of the present calculations (i.e. Comptonization
as the main process of high energy emission, uniform temperature, non-uniform
density) the time dependence of the photon flux, or light curves, at 
various energies can be used to map the radial density distribution of 
the hot electron Comptonizing cloud. This fact provides 
the possibility of deconvolution of the density structure of 
these clouds through timing analysis of their light curves.

\section{Conclusions}

1) For a given total Thompson depth of the Comptonization cloud, 
$\tau_0$, the electron density distribution can significantly
affect the emergent spectrum. Consequently, clouds of the same depth
$\tau_0$ and electron temperature $T_e$ can lead to different 
Comptonization spectra depending on their density distributions. On the 
other hand, clouds with different optical depths or temperature can lead 
to the same photon spectrum as long as their density profiles are properly
chosen. In other words, analysis of energy spectra alone cannot determine
the optical depth $\tau_0$ of the cloud, if its density structure is
unknown. Consequently, all the $\tau_0$ and $T_e$ values obtained by 
the traditional way of fitting analytically or numerically
calculated energy spectra with observational data are not the true
reflection of these properties but only indicate the effective values 
of a uniform plasma cloud.

2) The phase or time differences between X-rays in two energy bands
as functions of the Fourier frequency or period can provide additional, 
independent diagnostics of the light curve shapes and thereby of the density
structure of the scattering medium. This is because the phase or time
differences reflect the shape of the light curves, which in turn map
the density profiles of the clouds. In the framework of Comptonization,
the phase or time lags depend only on the physical size and density of 
the scattering medium. Therefore the information they provide is 
independent of possible alternative influences other than Compton 
scattering; such influences would generally afflict  the most common test of 
variability, namely the PSD. For example, in  timing analysis of the light
curves in terms of shots, the corresponding PSDs depend on the shot 
distribution, morphology and modulation of their rate of injection. The
time lag analysis is immune of all these influences. 

Therefore, combined analysis of spectral and temporal properties can 
yield knowledge not only about optical depth and plasma temperature but 
also about the physical size and density structure of the plasma clouds. 
When simultaneous observation of energy spectra and time variability,
especially the time lags, from a source are available, fitting of data in
the combined spectral and temporal domains such as those presented in this 
study, will provide less ambiguous information about these physical
properties that are important in the understanding of the the dynamics
of mass accretion onto the compact object. 

The authors would like to thank J.C. Ling, J.E. Grove, S.N. Zhang and
W. Focke for useful discussions.

\clearpage
\end{document}